\documentclass{article}
\usepackage{spconf,amsmath,graphicx}
\usepackage{amsfonts}
\usepackage{hyperref}
\usepackage{color}
\usepackage{booktabs}  
\usepackage[T1]{fontenc}

\title{A Phoneme-informed Neural Network Model for Note-level Singing Transcription}
\name{$^{1}$Sangeon Yong, $^{2}$Li Su, $^{1}$Juhan Nam}
\address{$^{1}$Graduate School of Culture Technology, KAIST, Daejeon, Republic of Korea\\$^{2}$Institute of Information Science, Academia Sinica, Taipei, Taiwan}
\begin{document}
\ninept
\maketitle
\begin{abstract}
Note-level automatic music transcription is one of the most representative music information retrieval (MIR) tasks and has been studied for various instruments to understand music.
However, due to the lack of high-quality labeled data, transcription of many instruments is still a challenging task.
In particular, in the case of singing, it is difficult to find accurate notes due to its expressiveness in pitch, timbre, and dynamics.
In this paper, we propose a method of finding note onsets of singing voice more accurately by leveraging
the linguistic characteristics of singing, which are not seen in other instruments.
The proposed model uses mel-scaled spectrogram and phonetic posteriorgram (PPG), a frame-wise likelihood of phoneme, as an input of the onset detection network while PPG is generated by the pre-trained network with singing and speech data.
To verify how linguistic features affect onset detection, we compare the evaluation results through the dataset with different languages and divide onset types for detailed analysis.
Our approach substantially improves the performance of singing transcription and therefore emphasizes the importance of linguistic features in singing analysis.
\end{abstract}
\begin{keywords}
singing transcription, onset detection, phoneme classification, music information retrieval
\end{keywords}

\section{Introduction}
\label{sec:intro}

Note-level singing transcription is an music information retrieval (MIR) task that predicts attributes of note events (i.e., onset time, offset time, and pitch value) from audio recordings of singing voice.
Although this task has been studied for a long time, the performance of singing transcription is generally inferior to those of other musical instruments such as polyphonic piano music \cite{onf, kong:21}.
The lack of large-scale labeled datasets is one of the major technical barriers. 
In addition, singing voice has highly diverse expressiveness in terms of pitch, timbre, dynamics, as well as phonation of lyrics. 
For example, singing techniques such as vibrato, bending, and portamento make it difficult to find note boundaries and note-level pitches.
This variability makes even manual note transcription by human experts difficult \cite{molina:14}. This in turn has resulted in the lack of high-quality labeled datasets.

Another important characteristic of singing voice which is well distinguished from other instruments is that it conveys linguistic information through lyrics and this influences note segmentation.
Given that most singing notes are syllabic (i.e., one syllable of text is set to one note of music) and melismatic (i.e., one syllable is sung with multiple notes), the relationship between the change of syllables and the change of notes is sophisticated.
This makes certain kinds of note patterns of singing voice not seen in any other instruments.
Therefore, we need to consider such linguistic characteristic in automatic singing transcription models.

\begin{figure}[!t]
\begin{minipage}[b]{1.0\linewidth}
  \centering
  \centerline{\includegraphics[width=\textwidth]{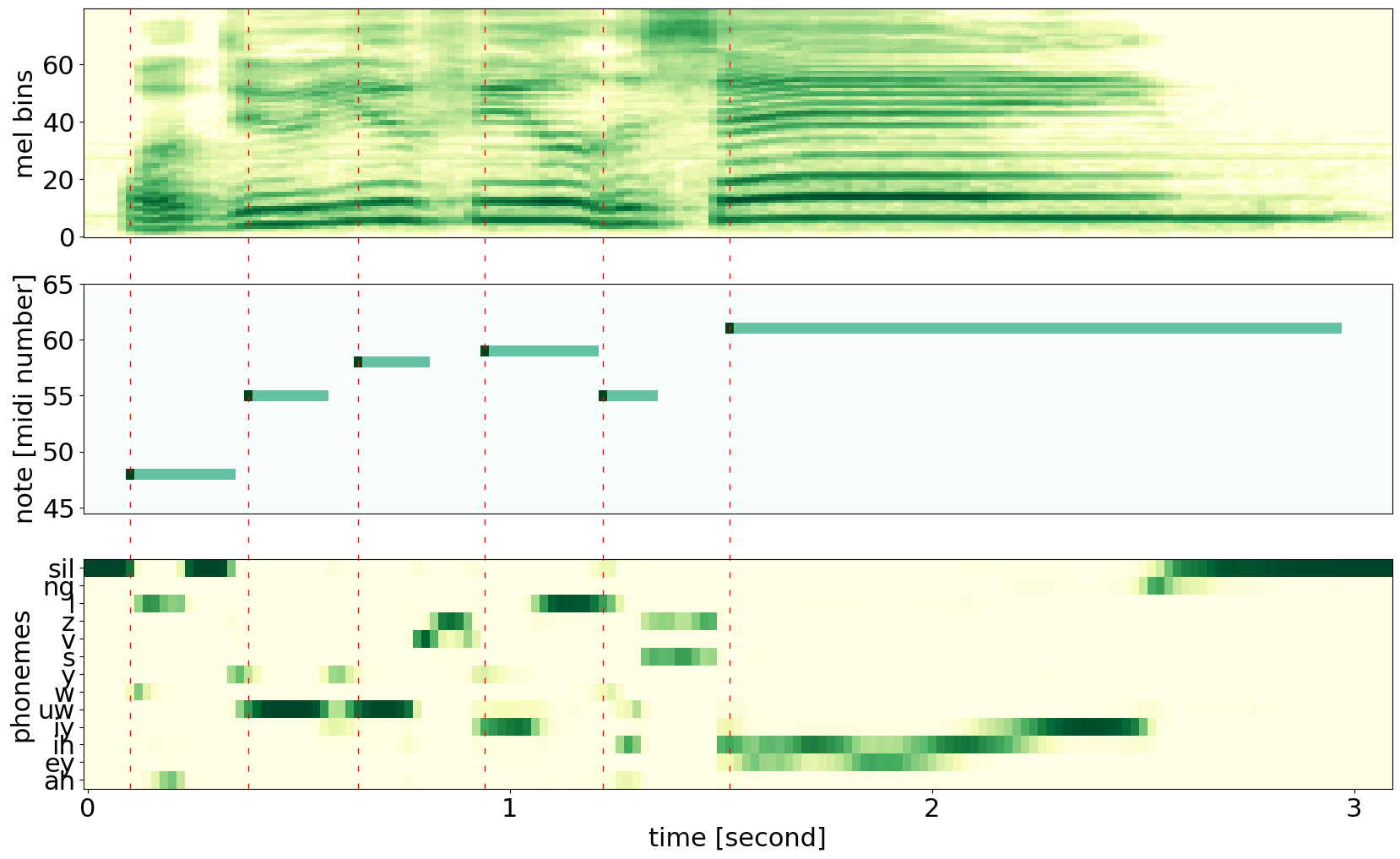}}
\end{minipage}
\caption{An example of singing voice: mel-spectrogram (top), piano roll with onsets and pitches of notes (middle), and phonetic posteriorgram (PPG) (bottom) from singing (phonemes with probability under 0.5 in this example were omitted).
}
\label{fig:phn_ex}
\vspace{-2mm}
\end{figure}

In this paper, we propose a neural network model that incorporates linguistic information into the input to improve note-level singing transcription for singing voice.
Similar to earlier research, we use log-scaled mel-spectrogram as a primary input. In addition to that, we take phonetic posteriorgram (PPG) from a pre-trained phoneme classifier as the second input.
As shown in Figure \ref{fig:phn_ex}, PPG shows a pattern distinct from the ones of mel-spectrogram, and it can be noted that the transition pattern of PPG can better describe the onset event at 1.2 and 2.8 second.
We propose a two-branch neural network model based on a convolutional recurrent neural network (CRNN) backbone to represent both of the input features effectively.
In the experiment, we conduct an ablation study to examine the effectiveness of model design, mel-spectrogram, and PPG.
Also, we compare the effects of mel-spectrogram and PPG on transition and re-onset, the two types of challenging onset events in singing transcription.
Finally, we demonstrate that our proposed model outperforms a few state-of-the-art note-level singing transcription models, especially in terms of onset detection.

\begin{figure}[!t]
\begin{minipage}[b]{1.0\linewidth}
  \centering
\centerline{\includegraphics[width=7.5cm]{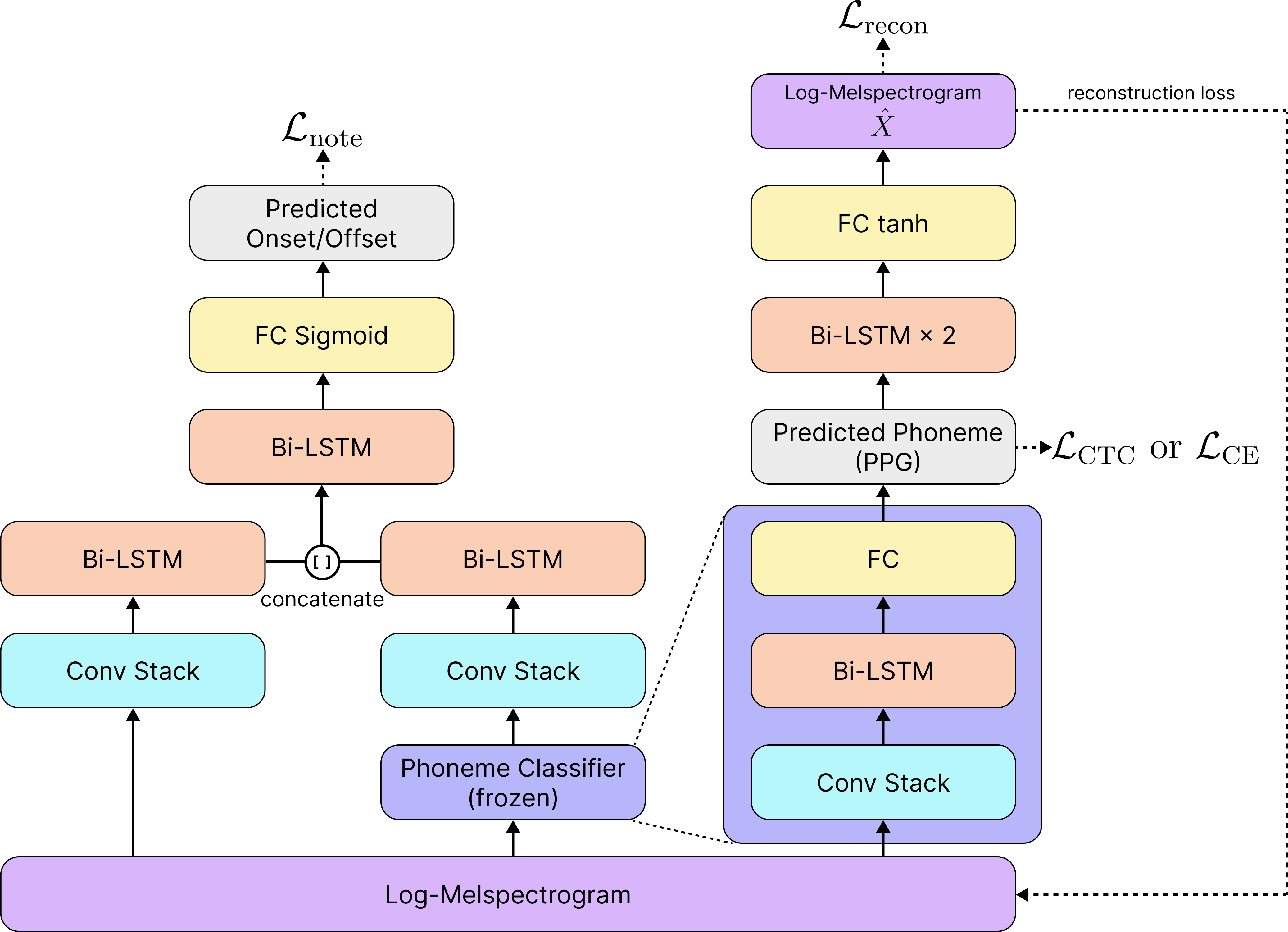}}
\end{minipage}
\caption{The proposed model architecture} 

\vspace{-3mm}

\label{fig:models}
\end{figure}

\section{Related Works}
\label{sec:related_works}

Traditional studies mainly used various types of spectral difference for onset detection of audio signals \cite{dixon:06}. 
The spectral difference is particularly successful at finding percussive onsets but it performs poorly on expressive instruments that have soft onsets.
Deep neural networks have been actively applied to singing voice as well.
Nishikimi \emph{et al}. \cite{nishikimi:19} suggested an attention-based encoder-decoder network with long short-term memory (LSTM) modules.
Fu \emph{et al}. \cite{fu:19} proposed a hierarchical structure of note change states to segment singing notes and used multi-channel features to increase the performance.
Hsu et al. \cite{hsu:21} suggested a semi-supervised AST framework.
More recently, \cite{wang:22} proposed the object detection-based approach to significantly improve the performance of singing voice onset/offset detection.
While the majority of them relied on note onset and offset information from melody labels, one recent attempted to use phoneme information as part of input features for note segmentation \cite{li:21}. However, the performance was not convincing. In this work, we present a neural network architecture to make an effective use of the phoneme information.



\begin{table*}[ht]
\footnotesize
\center
\begin{center}
\begin{tabular} {l|c|cccc|cccccc}
\toprule
Training dataset & & \multicolumn{4}{c}{SSVD v2.0}  & \multicolumn{6}{c}{CSD-refined} \\
Evaluation dataset & & \multicolumn{2}{c}{ISMIR2014} & \multicolumn{2}{c}{SSVD v2.0} & \multicolumn{2}{c}{CSD-refined} & \multicolumn{2}{c}{ISMIR2014} & \multicolumn{2}{c}{SSVD v2.0} \\
\midrule
& Feature & COn & COff & COn & COff & COn & COff & COn & COff & COn & COff \\
\midrule
(a) Single CRNN & $X$ & 0.8244 & 0.7751 & 0.8956 & 0.8983 & 0.9797 & 0.9719 & 0.8812 & 0.7524 & 0.8866 & 0.8007 \\
(b) Dual CRNNs + one RNN & $X,X$ & 0.9133 & 0.8513 & 0.9486 & 0.9566 & 0.9888 & 0.9838 & 0.9004 & 0.7636 & 0.8988 & 0.8089 \\
\midrule
(c) Single CRNN & $\hat{P}$ & 0.8655 & 0.7776 & 0.9223 & 0.9105 & 0.9890 & 0.9660 & 0.9048 & 0.7685 & 0.9063 & 0.8296 \\
(d) Dual CRNNs + one RNN & $\hat{P},\hat{P}$ & 0.9094 & 0.8310 & 0.9342 & 0.9470 & 0.9907 & 0.9638 & 0.9090 & 0.7733 & 0.9142 & 0.8336 \\
\midrule
(e) Dual CNNs + one RNN & $X, \hat{P}$ & 0.9024 & 0.8349 & 0.9439 & 0.9420 & 0.9877 & 0.9791 & 0.9016 & 0.7852 & 0.9098 & \textbf{0.8340} \\
(f) Dual CNNs + two RNNs & $X, \hat{P}$ & 0.9230 & 0.8538 & 0.9496 & 0.9531 & 0.9914 & 0.9839 & \textbf{0.9150} & \textbf{0.7804} & \textbf{0.9199} & 0.8328 \\
(g) Dual CRNNs + one RNN & $X, \hat{P}$ & \textbf{0.9305} & \textbf{0.8576} & \textbf{0.9569} & \textbf{0.9692} & \textbf{0.9923} & \textbf{0.9864} & 0.9145 & 0.7723 & 0.9166 & 0.8257 \\

\bottomrule
\end{tabular}
\end{center}
\caption{Onset/Offset detection results from various neural network architectures with two input features. $X$ and $\hat{P}$ denote mel-spectrogram and PPG, respectively. (g) corresponds to the neural network architecture in Figure 2. }

\label{table:ablation_study}
\end{table*}

\section{Proposed Method}\label{sec:method}


\subsection{Model Architecture}

Our proposed model architecture consists of two branch networks and a single RNN with a dense layer as illustrated in Figure \ref{fig:models}.
One branch network takes log-scaled mel-spectrogram $X$ and the other branch network takes phonetic posteriorgram (PPG) $\hat{P}$ from a pretrained phoneme classifier.
Both of the branches are CRNN where CNN architectures are a modified version of \emph{ConvNet} proposed in \cite{kelz:16}, which is commonly used in the piano transcription task \cite{onf, kwon:20}.
To get the wider time-scale receptive field, we changed the first convolution layer with a dilated convolution with 2 dilation on the time frame axis.
To predict the note events, we combined the two branch networks by concatenating the outputs and connecting them to an additional RNN layer and a dense layer. The output layer is represented with a 3-dimensional sigmoid vector where each element detects onset, offset, and activation as binary states. The activation indicates whether the note is on or off at each frame.   


\subsection{Framewise Phoneme Classifier}\label{subsec:phoneme_classifier}

We extracted the phonetic information using a phoneme classifier which returns the output as a PPG. We implemented it using a single CRNN network with a dense layer. 
We used the original \emph{ConvNet} architecture for the CNN part.
We tried two loss functions to train the phoneme classifier network. One is the framewise cross entropy loss, which is possible when we have time-aligned phoneme labels. 
Since it is difficult to obtain time-aligned phoneme labels in frame-level especially for singing voice, we also used the connectionist temporal classification (CTC) loss function \cite{graves:06} which can handle the alignment between the predicted phoneme sequence ($\hat{p}$) and the ground truth phoneme sequence ($S$) which have unequal lengths.
The CTC algorithm predicts phoneme sequences with inserted blank labels along the possible prediction paths $\mathcal{B}$.
Since the CTC loss function is optimized for predicting the entire sequence,
the prediction pattern tends to be spiky and sparse and thus it does not find the boundaries of phonemes well \cite{graves:06, sak:15}.
To solve this problem, we used two layers of bidirectional LSTM layers and a single dense layer that reconstruct the input log-scaled mel-spectrogram ($\hat{X}$).
This was proposed to enhance the time alignment when the CTC loss is used \cite{teytaut:21}.
For the reconstruction loss ($\mathcal{L}_{\text{recon}}$), we normalized the log-scaled mel-spectrogram from $-1$ to $1$ ($\Tilde{X}$) and applied the $\tanh$ function for the activation and used the $L_2$ loss function.
These loss functions are defined as:

\vspace{-2mm}
\begin{align}
    \mathcal{L}_{\text{CTC}} &= -\log\sum_{\hat{p}, \mathcal{B}(\hat{p})=p} \prod_{t=0}^{T-1} \mathbb{P}(\hat{p}_{t}|X)\,, \nonumber \\
\vspace{2mm}
    \mathcal{L}_{\text{recon}} &= \| \hat{X} - \Tilde{X} \|^{2}\,, \\
\vspace{2mm}
    \mathcal{L}_{\text{PPG}} &= \mathcal{L}_{\text{CTC}} + \mathcal{L}_{\text{recon}}\,, \nonumber
    \label{phoneme_classifier_loss}
\end{align}
where $T$ is the total number of time steps, $p$ is the ground truth phoneme sequence and $\mathbb{P}(\hat{p}_{t}|X)$ is the PPG at time $t$. 

\subsection{Label Smoothing}

Unlike other instruments, synthesized or auto-aligned onset/offset labels are hardly available in the case of the singing datasets \cite{maestro}.
In addition, since singing onsets are temporally soft, has a soft onset, to locate the exact onset positions of singing by means of with a waveform or mel-spectrogram is by no means straightforward.
Such softness of the onset is one of the factors that makes the onset of singing voices more challenging to train.
Previous frame-wise onset detection studies \cite{fu:19, hsu:21} extended the duration of the onset label to solve this problem.

Following these previous studies, we also used a smoothing method to increase the length of the onset and offset label. 
Specifically, we smoothed the 1-D one-hot onset label sequence $y_{\text{on}}:=y_{\text{on}}[n]$ ($n$ denotes the time index) and the offset label sequence $y_{\text{off}}:=y_{\text{off}}[n]$ through the linear convolution with a scaled triangular window function $w_{\text{tri}}[n]$ to improve the precision simultaneously. The scale factor of the triangular function $N$ stands for the number of frames with nonzero values. To make the center of the label to $1$ after the smoothing, we only used the odd numbers for the scale factor $N$. The convolution process is represented as
\begin{align}
  w[n] &=
    \begin{cases}
      1 - \left|\frac{n}{(N + 1) / 2}\right| & \text{if $|n| \leq \frac{(N + 1)}{2}$}  \\
      0 & \text{otherwise.} 
    \end{cases} \nonumber \\
  y_{\text{on\_s}}[n] &= y_{\text{on}}[n] \ast w_{\text{tri}}[n] \\
  y_{\text{off\_s}}[n] &= y_{\text{off}}[n] \ast w_{\text{tri}}[n] \nonumber
  \label{eq:label_smoothing}
\end{align}
where the operation $\ast$ represents the linear convolution and $n$ is the frame index. 

\subsection{Note Decoding}\label{subsec:label_representation}

To find the positions of onsets from the prediction output, we set a constant threshold and set the frame with the maximal value above the threshold as the position of onset.
When finding the offset of a note, we first find the offset candidates between the current onset time and the next onset time. The offset candidate is either the highest peak of the offset prediction or the time frame that the activation prediction goes lower than 0.5. If multiple offset candidates exist, we set the offset to the latest offset candidate. If no offset candidate is found, the offset of the note is set to the time frame of the next onset. The threshold of onset and offset is set to 0.2. In order to determine the threshold, we evaluated the validation set using a threshold ranging from 0.1 to 0.9 in increments of 0.1 to identify the optimal threshold.

For note-level singing transcription, we estimated the note-level pitch from frame-wise F0s of the note segment to find the pitch of the note, following \cite{fu:19}. 
We extracted F0s with the PYIN algorithm \cite{pyin}, which is one of the most accurate pitch trackers. To compress the F0 contour to the note-level pitch, we used the weighted median algorithm, which finds the 50\% percentile in the ordered elements with given weights.  In this experiment, we use the normalized Hann window function with the same length of the note segment frames as the weight of the weighted median to reduce the influence of the F0 near the boundaries, which are the most expressive part. Since the sum of all weight values should be one, the Hann window function is normalized by dividing by the sum of the window elements.

\section{Experiments}\label{sec:experiments}

\subsection{Datasets}

We used SSVD v2.0 as the primary dataset \cite{wang:22}. It contains multiple sight-singing recordings, consisting of 67 singing audio files for the train and validation set, and 127 audio files for the test set. The human labeled annotations include onset, offset, and averaged note pitch. To use both phoneme and note labels given the audio, we also used the 50 songs in Korean from the CSD dataset \cite{CSD}, which have both note and phoneme labels of a female professional singer. Since the original note annotations of CSD was targeted for singing voice synthesis, we found it needs some refinement for the note transcription task. Thus, we re-annotated 50 songs of CSD for our experiment, following the rule suggested by \cite{molina:14}. The re-annotated label of CSD can be found on our GitHub page \footnote{\href{https://github.com/seyong92/CSD_reannotation}{https://github.com/seyong92/CSD\_reannotation}}.
The refined CSD is split 35, 5, and 10 songs for train, validation, and test set each.

To train the phoneme classifier, we used TIMIT \cite{TIMIT} which contains English speech with time-aligned phoneme labels for the model with SSVD v2.0. TIMIT contains 5.4 hours of audio of English speech. While training the phoneme classifier network, we reduced the phoneme types to 39 following the CMU pronouncing dictionary \cite{cmudict}. For the model with CSD, we used the unaligned phoneme label in CSD to train.

To compare the transcription performance of the proposed model with previous work, we also used the ISMIR2014 \cite{molina:14} dataset, which contains 38 songs sung by both adults and children, as a test set.

\vspace{-3mm}
\subsection{Evaluation Metrics}

We evaluated the models with the \texttt{mir\_eval} library \cite{mir_eval} for onset/offset detection and note-level transcription.
We used the metrics proposed in \cite{molina:14}: F1-measure of COn (correct onset), COff (correct offset), COnOff (correct onset and offset), COnP (correct onset and pitch), and COnPOff (Correct onset, offset and pitch).
We used the default parameters of \texttt{mir\_eval}, which sets the onset tolerance to 50 ms, the offset tolerance to larger value between 50 ms and 0.2 of note duration, and the 50 cents for the pitch tolerance. Also, we report the results when the onset/off thresholds are 100 ms considering the softness of singing onsets.

\subsection{Training Details}
We computed 80 bin mel-spectrogram $X$ with 320 samples in hop size (20 ms) and 1024 samples in FFT size after resampling audio files to 16 kHz. For the modified \emph{ConvNet} module, we set 48/48/96 nodes to the convolutional layers and 768 nodes to the dense layer. We used 768 nodes in all bidirectional LSTM layers and set the last FC layer in the note onset/activation detector to have two separate nodes for onset and activation detection, respectively. For the label smoothing, we used a scale factor of 5 to extend the label length to 100 ms, which shows the best results in our experiment.

To train the note onset/offset detection network, we used the AdamW optimizer \cite{loshchilov:19} with a batch size of 8 and a learning rate of 1e-6. We reduced the learning rate with a reducing factor of 0.98 for every 1000 steps. While training, we used the random audio segment with 5 seconds. The validation set was evaluated for every 500 steps and we stopped training when there is no advance in the model for 10 validation steps. To train the phoneme classifier, we used the Adam optimizer with a batch size of 16 and a learning rate of 2e-4. We reduced the learning rate with a reducing factor of 0.98 for every 900 steps. We validated the model with every 500 steps for the phoneme classifier and trained the model while there is no advance in the model for 5 validation steps.

\begin{figure}[!t]
\begin{minipage}[b]{1.0\linewidth}
  \centering
 \vspace{-2mm}
\centerline{\includegraphics[width=6.5cm]{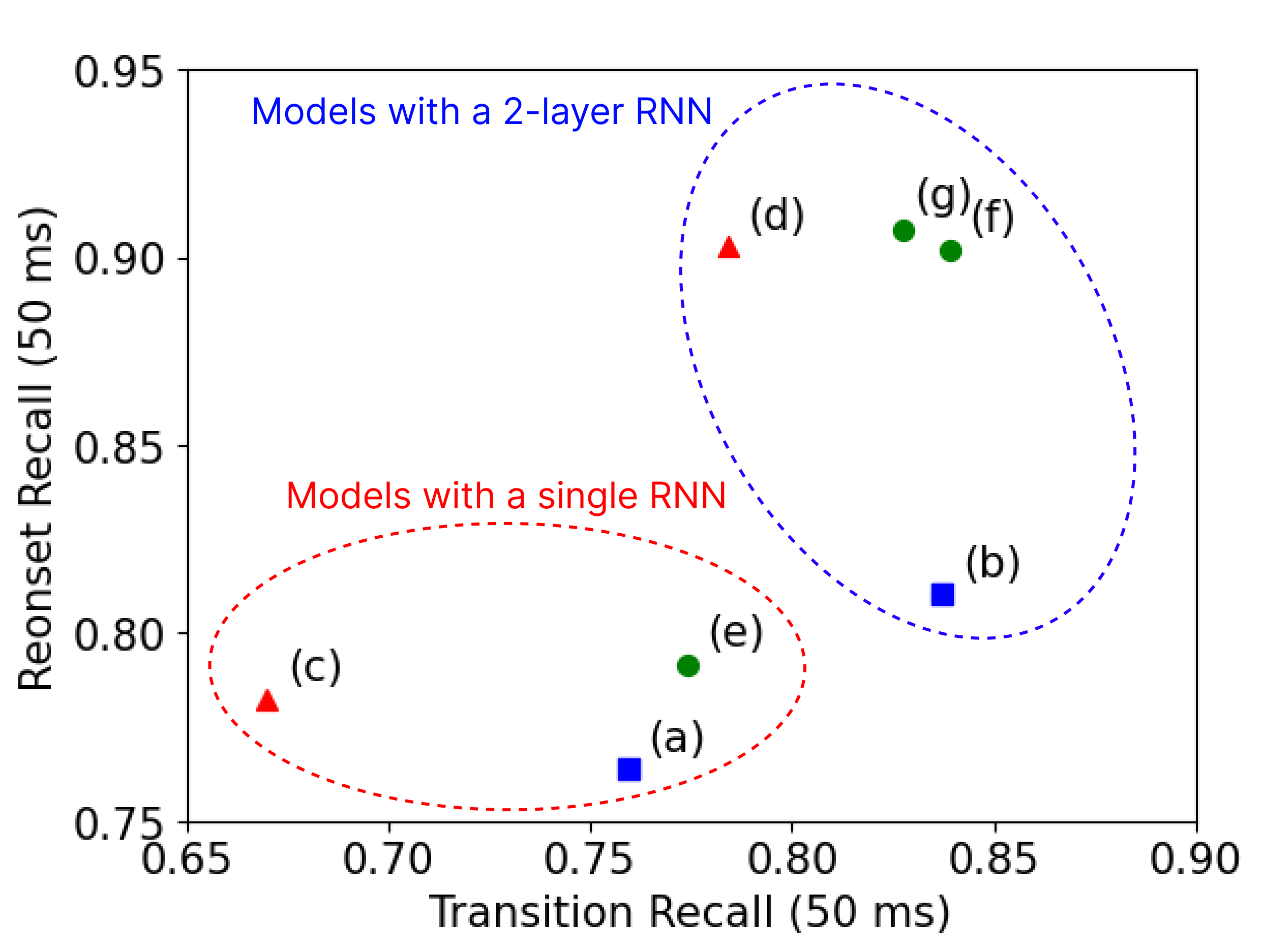}}
\end{minipage}
\vspace{-4mm}
\caption{Transition and re-onset recall of the models in the ablation study on ISMIR2014. The red triangle is the model with mel-spectrogram, the blue square is the model with PPG, and the green circle is the model with both features.}
\label{fig:onset_type_analysis}
\end{figure}

\begin{table*}[ht]
\small
\renewcommand{\tabcolsep}{0.12cm}
\center
\begin{center}
\begin{tabular} {ccccccccccccc}
\toprule
Model & \multicolumn{3}{c}{COn (50ms)} & \multicolumn{3}{c}{COn (100ms)} & \multicolumn{3}{c}{COff (50ms)} & \multicolumn{3}{c}{COff (100ms)} \\
& P & R & F & P & R & F & P & R & F & P & R & F \\
\midrule
TONY \cite{mauch:15} & 0.7068 & 0.6326 & 0.6645 & 0.8402 & 0.7486 & 0.7877 & 0.7862 & 0.6981 & 0.7358 & 0.8405 & 0.7471 & 0.7870 \\
Omnizart \cite{hsu:21, omnizart} & 0.7797 & 0.8229 & 0.7951 & 0.8667 & 0.9153 & 0.8843 & 0.7698 & 0.8132 & 0.7852 & 0.8394 & 0.8842 & 0.8554 \\
MusicYOLO (retrained) \cite{wang:22} & 0.9427 & 0.8970 & 0.9176 & \textbf{0.9711} & 0.9247 & 0.9456 & \textbf{0.8924} & \textbf{0.8504} & \textbf{0.8693} & \textbf{0.9476} & 0.9024 & 0.9227 \\
\textbf{Proposed} & \textbf{0.9448} & \textbf{0.9188} & \textbf{0.9305} & 0.9652 & \textbf{0.9387} & \textbf{0.9506} & 0.8701 & 0.8473 & 0.8576 & 0.9429 & \textbf{0.9176} & \textbf{0.9290} \\
\bottomrule
\end{tabular}
\end{center}
\vspace{-2mm}
\caption{Onset/Offset detection results on ISMIR2014. Both of MusicYOLO and the proposed model were trained with SSVD v2.0. Omnizart is a pretrained note transcription model package (not with SSVD v2.0). Tony is a free, open-source application for pitch and note transcription.}
\label{table:ismir2014_eval}
\end{table*}

\vspace{-3mm}
\section{Results and Discussions}\label{sec:results}

\subsection{Ablation Study}
We conducted an ablation study to see the effect of input features and model architectures. The proposed model shown in Figure \ref{fig:models} corresponds to "Dual CRNNs + one RNN" in (g). We first compare it to a single CRNN model with only one type of features (either mel spectrogram in (a) or PPG in (c)). Considering that the model architecture can affect the performance, we also compared the proposed model to the same "Dual CRNNs + one RNN" but with one type of input features for both inputs (either mel spectrogram in (b) or PPG in (d)). Given the proposed model, we also removed the RNN module in each CRNN branch in (e), and then stacked another RNN module on top of (e) in (f).    

Table \ref{table:ablation_study} show the onset/offset detection results of all compared models. Single CRNNs with only one input features in (a) and (c) have significantly lower accuracy than the proposed model in (g). The gap is relatively lower when the model was trained with CSD. Interestingly, the single CRNN model with PPG consistently outperformed the one with mel spectrogram. The results from the same model architecture with different input features in (b), (d), and (g) shows that using both mel-spectrogram and PPG is more effective than using either one of them. However, the gaps are less significant than those in the comparison with single CRNN in (a) and (c). This indicates that model architecture is also important to improve the performance. Likewise, the results in (e), (f), and (g) show that the design choice of neural network affects the performance. Since CSD is a small dataset, the proposed model have a tendency to overfit it. Overall, the propose model in (g) shows the best performance.  

We further investigated the effect of the input features by looking into the recall accuracy for two special types of onsets: re-onset and transition. They are note onsets which have 20 ms or less apart from the offset of the previous note. The difference between the two types is whether the pitch changes (transition) or not (re-onset). The re-onset usually occurs when the syllable in lyrics or energy changes while continuing the same pitch. Note that, since our model does not predict the onset types, only recall accuracy can be computed. As shown in Figure \ref{fig:onset_type_analysis}, the models with mel-spectrogram (in red) tend to detect more transitions, indicating that it is more sensitive to pitch change. On the other hand, the models with PPG (in blue) tend to detect more re-onsets, showing that it captures phonetic changes well. Lastly, the models with both features have more balanced accuracy in both transition and re-onset. The demo examples, more analysis, and pre-trained models are available on the companion website. \footnote{\href{https://seyong92.github.io/phoneme-informed-transcription-blog/}{https://seyong92.github.io/phoneme-informed-transcription-blog/}}

\subsection{Comparison with Prior Work}
Table \ref{table:ismir2014_eval} shows the comparison with prior work on the ISMIR2014 dataset, which has been widely used for singing voice onset/offset detection (or note segmentation). For fair comparison, we retrained a recent state-of-the-art model \cite{wang:22} with the same dataset we used for the proposed model. Our proposed model outperforms the state-of-the-art model in onset F-score in both tolerances while it is slightly worse in offset F-score in 50ms tolerance. The publicly available note transcription software (TONY) and model package (Omnizart) have significantly lower accuracy than the two models. Finally, to see the performance for singing note transcription including pitch information, we measured COnP and COnPOff on ISMIR2014 and SSVD v2.0 in Table \ref{table:ismir2014_ssvd_note_eval}. 
The results show that the proposed model achieves consistently better performances than TONY and Omnizart. 

\begin{table}[!t]
\small
\center
\begin{center}
\begin{tabular} {ccccc}
\toprule
& \multicolumn{2}{c}{ISMIR2014} & \multicolumn{2}{c}{SSVD v2.0} \\
\midrule
Model & COnP & COnPOff & COnP & COnPOff \\
\midrule
Tony \cite{mauch:15} & 0.6009 & 0.4621 & 0.7311 & 0.6794 \\
Omnizart \cite{hsu:21, omnizart} & 0.6174 & 0.4992 & 0.6047 & 0.5151 \\
\textbf{Proposed} & \textbf{0.8975} & \textbf{0.7728} & \textbf{0.8558} & \textbf{0.8303} \\

\bottomrule
\end{tabular}
\end{center}
\vspace{-2mm}
\caption{Note transcription results on ISMIR2014 and SSVD v2.0. The proposed model was trained with SSVD v2.0 }

\label{table:ismir2014_ssvd_note_eval}
\end{table}

\vspace{-3mm}
\section{Conclusion}\label{sec:conclusion}
We presented a neural network architecture for note-level singing transcription that takes advantage of PPG on top of mel-spectrogram. Through the ablation study, we examined various architectures along with the two input features, showing that the additional phonetic information is effective in singing onset/offset detection. Also, we showed that the proposed model outperforms the compared models on ISMIR2014 and SSVD v2.0. For future work, we plan to explore models that effectively handle weak supervision from noisy melody and lyrics labels on a large-scaled dataset \cite{dali}. 

\bibliographystyle{IEEEbib}
\bibliography{strings,refs}

\end{document}